# Analytical Representation for the Electronic Contribution of the Nuclear Schiff Interaction Hamiltonian

Satoshi Toda[a], Yasuto Masuda[a], Naohiro Tomiyama[b], Kota Yanase[c], Bijaya Kumar Sahoo[d], Masahiko Hada[b], and Minori Abe[a,e]*

[a]*Department of Chemistry, Hiroshima University, Higashi-Hiroshima City, Japan;*
[b] *Department of Chemistry, Tokyo Metropolitan University, Hachioji City, Japan;*
[c] *Nishina Center for Accelerator-Based Science, RIKEN, Wako, Japan;*
[d]*Atomic, Molecular and Optical Physics Division, Physical Research Laboratory, Ahmedabad, India;*
[e]*Department of Applied Physics and Chemical Engineering, Tokyo University of Agriculture and Technology, Koganei, Japan*

*Minori Abe, Department of Applied Physics and Chemical Engineering, Tokyo University of Agriculture and Technology, Koganei, Tokyo 184-8588, Japan, minoria@go.tuat.ac.jp

# Analytical Representation for the Electronic Contribution of the Nuclear Schiff Interaction Hamiltonian

The nuclear Schiff interaction (NSI) arises from a nuclear force that simultaneously violates spatial parity (P) and time reversal (T) symmetries, where T symmetry is equivalent to CP symmetry under CPT invariance. Detecting the NSI experimentally is important because CP violation is critical for explaining why the amount of matter in the Universe is far greater than that of antimatter. Measuring the NSI in molecules requires both precise experiments and theoretical calculations that incorporate electronic and nuclear wavefunctions. Conventionally, the electronic terms have been approximated using a first-order power series expansion of the electronic radial function—an approach that yields the well-known nuclear Schiff moment (NSM) —but this approximation may not be sufficiently accurate. In this study, we introduce a new, accurate analytical expression for the electronic terms based on Gaussian basis sets, which avoids any truncation of the power series. We find that the previous numerical approach overestimates the values for RaO and LrF by more than 50% and 300%, respectively, in the nuclear-radius region. In contrast to the numerical calculations, the analytical expression-based calculations show less sensitivity to choice of the basis-functions. Furthermore, we develop a new basis set that describes accurate behavior of wave functions both interior and exterior regions of nucleus. It also demonstrates that an even-tempered basis set is more preferrable over energy optimized basis set for calculating the NSI electronic term in molecules.

## Introduction

At the time of the Big Bang, equal numbers of particles and antiparticles were produced, yet the Universe is now dominated by matter. Although this matter–antimatter asymmetry is generally attributed to the violation of charge–parity (CP) symmetry [1], the observed level of CP violation is insufficient to quantitatively account for the current state of the Universe, making the discovery of larger CP-violating effects essential [2, 3]. In this context, measurements of the electric dipole moment (EDM) of fundamental particles have attracted considerable attention as potential evidence for new CP-violating phenomena[4]. In particular, experiments involving molecules containing heavy atoms can offer enhanced sensitivity compared to those using atoms alone [5,6]. Consequently, recent electron EDM experiments have been actively conducted using molecules such as YbF [7], ThO [8,9], and $HfF^+$ [10].

In this study, we focus on the nuclear Schiff interaction (NSI) [11], one of the primary contributors to atomic and molecular EDMs in closed-shell electronic systems, where contributions from the electron EDM are absent. For NSI, experimental upper bounds have been reported for the Ra[12,13], Hg [14], Xe [15], and TlF [16] systems. The NSI originates from an asymmetric charge distribution within the nucleus, induced by nucleon–nucleon interactions that violate both parity (P) and time-reversal (T) symmetries (noting that T violation is equivalent to CP violation under the CPT theorem) [17]. Determining the NSI requires not only high-sensitivity experimental measurements but also rigorous theoretical calculations, including those based on nuclear wave functions [18,19] and relativistic electronic structure theory to evaluate the electronic contributions [13,20-24].

Conventionally, the electronic contribution has been approximated by the derivative of the electron density at the nuclear origin, $\left(\nabla\psi^{\dagger}\,\psi\right)_{r=0}$. Hereafter, we refer to this approach as the *conventional* method. In this context, the well-known nuclear Schiff moment (NSM) is given as follows [25].

$$\mathbf{S} = \frac{e}{10}\left[\left(\int_0^{\infty} \mathbf{r_n} r_n^2 \rho(\mathbf{r_n}) d^3\mathbf{r_n}\right) - \frac{5}{3Z}\left(\int_0^{\infty} \mathbf{r_n} d^3\mathbf{r_n}\right)\cdot\left(\int_0^{\infty} \rho(\mathbf{r_n}) d^3\mathbf{r_n}\right)\right]$$

which can be written in terms of expectation values as

$$\frac{e}{10}\left[\langle \mathbf{r_n} {r_n}^2\rangle - \frac{5}{3Z}\langle \mathbf{r_n}\rangle\langle {r_n}^2\rangle\right] = S\mathbf{I}/I. \qquad (1)$$

Here, $\mathbf{r_n}$, $\rho(\mathbf{r_n})$, $Z$, and $\mathbf{I}$ denote nuclear coordinate, proton density, atomic number, nuclear spin, respectively. However, this approximation may be insufficiently accurate for heavy atoms, which have finite and large nuclear charge distributions. Therefore, we introduce a new analytical representation of the electronic contribution using an expansion in Gaussian basis set functions, which are commonly employed to describe electronic wave functions in molecules. Basis set dependence was investigated by comparing the results obtained with an even-tempered (ET) basis set and an energy-optimized basis set (Dyall.cv4z). At the mean-field approximation, we find that the ET basis set more accurately describes the electronic contribution to the NSI than Dyall.cv4z and that the error in the conventional approximation is significant in RaO and LrF molecules. Furthermore, by incorporating electron correlation effects up to the coupled-cluster singles and doubles (CCSD) level, we achieve a more precise calculation of the electronic contributions.

## 2. Theories

### *2.1 Nuclear Electrostatic Potential of the NSI*

Considering the electron shielding effects [26], the nuclear electrostatic potential $\varphi(\mathbf{r})$ can be expressed as follows [26-28].

$$\varphi(\mathbf{r}) = \int \frac{e\rho(\mathbf{r_n})}{|\mathbf{r}-\mathbf{r_n}|} d^3\mathbf{r_n} + \frac{1}{Z}(\mathbf{d}\cdot\boldsymbol{\nabla}) \int \frac{\rho(\mathbf{r_n})}{|\mathbf{r}-\mathbf{r_n}|} d^3\mathbf{r_n} \tag{2}$$

Here, $\boldsymbol{Z}$ denotes the atomic number, $\mathbf{r}$ represents the electronic coordinate, $\mathbf{r_n}$ the nuclear coordinates, and $\boldsymbol{\nabla}$ denotes the gradient with respect to the electronic coordinate $\mathbf{r}$. Furthermore, $e\rho(\mathbf{r_n})$ denotes the nuclear charge density, and $\mathbf{d}$ is the nuclear EDM, calculated as $\mathbf{d} = e\int \rho(\mathbf{r_n})\mathbf{r_n} d^3\mathbf{r_n} \equiv e\langle\mathbf{r_n}\rangle$. Since this study focuses on the NSI of diatomic molecules containing a heavy atom, we adopt a coordinate system in which the heavy atom is placed at the origin; consequently, $\varphi(\mathbf{r})$ denotes the nuclear electrostatic potential of that heavy atom. The electrostatic potential $\varphi(\mathbf{r})$ can be expanded in a multipole series using Legendre polynomials as:

$$\frac{1}{|r-r_n|} = \frac{r_<^l}{r_>^{l+1}} P_l(\cos\theta) \tag{3}$$

Here, $r_<$ denotes $min[r, r_n]$ , and $r_>$ denotes $max[r, r_n]$. In the multipole expansion, only terms with odd $l$ break P and T symmetry. Moreover, since contributions from terms with $l \geq 3$ (electric octapole and higher moments) are small [29], the CP-violating part of the electrostatic potential $\varphi(\mathbf{r})$ can be written by focusing on the $l = 1$ term (with the second term in Equation (2) corresponding to $l = 0$) as follows [5]:

$$\varphi^{(1)}(\mathbf{r}) = e\mathbf{r} \cdot \left[\int_r^\infty \left(\frac{\langle \mathbf{r_n} \rangle}{Zr^3} - \frac{\mathbf{r_n}}{r^3} + \frac{\mathbf{r_n}}{r_n^3}\right) \rho(\mathbf{r_n}) d^3\mathbf{r_n}\right], \tag{4}$$

where Equation (4) corresponds to the nuclear electrostatic potential arising from the NSI. Here, the integration variable $\mathbf{r_n}$ in Eq. (4) is expressed in spherical coordinates.

### *2.2 Interaction Energy*

In a one-electron system, the dominant contribution to the interaction energy between the electron and the NSI-induced electrostatic potential, $\varphi^{(1)}(\mathbf{r})$ [25], is given by the following matrix element:

$$\langle s| - e\varphi^{(1)}(\mathbf{r})|p\rangle = -e^2 \left\{\int_0^\infty \left[\frac{1}{Z}\langle \mathbf{r_n} \rangle \cdot \langle s|\boldsymbol{n}|p\rangle \int_0^{r_n} U_{sp}(r)dr + \mathbf{r_n} \cdot \langle s|\boldsymbol{n}|p\rangle \left(\frac{1}{r_n^3}\int_0^{r_n} U_{sp}(r)\, r^3 dr - \int_0^{r_n} U_{sp}(r)\, dr\right)\right] \rho(\mathbf{r_n}) d^3\mathbf{r_n}\right\} \tag{5}$$

Here, $\mathbf{n} = \mathbf{r}/r$, and $\langle s|n|p\rangle$ denotes an angular integral over the electronic coordinates $\theta$and $\phi$. Since $\varphi^{(1)}(\mathbf{r})$ is a parity-violating operator in the vicinity of the nucleus and has significant components only in that region, we focus exclusively on the matrix elements between s and p orbitals, which possess opposite parities. Other contributions arising from p–d and d-f interactions, together with contributions from light-element atomic orbitals, are neglected in the present work. $U_{sp}(r)$ is expressed as follows:

$$U_{sp}(r) = f_s(r)f_p(r) + g_s(r)g_p(r) \tag{6}$$

Here, $f(r)$ and $g(r)$ denote the radial functions of the large and small components of the atomic orbitals, respectively. In Equation (5), the two terms that depend on the electronic state become functions of the nuclear radial coordinate $r_n$; these are referred to as electronic state term 1, $F_1(r_n)$, and electronic state term 2, $F_2(r_n)$. The "$s_{1/2}$ and $p_{1/2}$" or "$s_{1/2}$ and $p_{3/2}$" orbital combinations are chosen to have the same $m_j=\pm1/2$, resulting in a total of four possible patterns. Furthermore, if the z-axis is chosen along the molecular axis, only the z-component of $\langle s|\mathbf{n}|p\rangle$ is nonzero, and each electronic term can be written as follows [25].

$$\langle s| - e\varphi^{(1)}(\mathbf{r})|p\rangle = -e^2 \int_0^\infty \left[\frac{1}{Z}\langle \mathbf{r_n}\rangle_z \cdot F_1(r_n) + (\mathbf{r_n})_z \cdot F_2(r_n)\right] \rho(\mathbf{r_n}) d^3\mathbf{r_n} \,, \tag{7}$$

$$F_1(r_n) = (\langle s|\boldsymbol{n}|p\rangle)_z \int_0^{r_n} U_{sp}(r)dr, \tag{8}$$

and

$$F_2(r_n) = (\langle s|\boldsymbol{n}|p\rangle)_z \left(\frac{1}{r_n^3}\int_0^{r_n} U_{sp}(r)r^3 dr - \int_0^{r_n} U_{sp}(r)dr\right). \tag{9}$$

Note that the electronic terms are computed by integrating $U_{sp}(r)$ and $U_{sp}(r)r^3$ over the range $0 \leq r \leq r_n$. Furthermore, since the interaction energy given in Equation (5) is obtained by integrating the product of the electronic term and $\rho(\mathbf{r_n})$ over the nuclear coordinate $\mathbf{r_n}$, it follows that the interaction energy is highly sensitive to the electronic state within the nucleus. For heavy atoms, the properties in the vicinity of the nucleus are significantly affected by relativistic effects. Therefore, to accurately evaluate the interaction energy, relativistic effects must be taken into account using either two-component or four-component relativistic methods. In the present work, we adopt a four-component relativistic formalism, and all equations are formulated accordingly.

## *2.3 Electronic Terms*

### *2.3.1 Electronic Term of Conventional (Point-like) Representation*

In the conventional approach, the radial functions, $f(r)$ and $g(r)$, are expanded in a Maclaurin series, and the lowest-order term is used to represent $U_{sp}(r)$. That is, an approximate expression for $U_{sp}(r)$ is employed as follows [25].

$$U_{sp}(r) = f_s(r)f_p(r) + g_s(r)g_p(r) = \sum_k \left(b_k^{large} + b_k^{small}\right) r^k = \sum_k b_k r^k \approx b_1 r \quad (10)$$

The electronic terms arising from the large and small components of the molecular orbitals, bra: $\phi_i^*$ and ket: $\phi_j$ , are denoted as $[F^{large}(r_n)]_{i,j}$ and $[F^{small}(r_n)]_{i,j}$ , respectively. Using Gaussian basis set expansion [30] in Equation (9), the contribution from the large component to electronic terms 1 and 2 for the $s_{1/2}$ ($m_j$=+1/2) and $p_{1/2}$ ($m_j$=+1/2) combination, with $b_1^{large} = 1,\ b_1^{small} = -12\alpha_k^*,$ and $(\langle s|\mathbf{n}|p\rangle)_z = 4\pi/3\sqrt{3}$ is given below. (Conventional expressions for other combinations are provided in appendices.) It yields

$$[F_1(r_n)]_{i,j} = \left[F_1{}^{large}(r_n)\right]_{i,j} + \left[F_1{}^{small}(r_n)\right]_{i,j}$$

,

with

$$\left[F_1{}^{large}(r_n)\right]_{i,j} = \sum_{k,l} C_{k,i}^{sL^*} C_{l,j}^{pL} N_k^{sL^*} N_l^{pL} (\langle s|\boldsymbol{n}|p\rangle)_z \int_0^{r_n} b_1^{large} r dr$$

$$= \sum_{k,l} C_{k,i}^{sL^*} C_{l,j}^{pL} N_k^{sL^*} N_l^{pL} \frac{2\pi}{3\sqrt{3}} r_n{}^2$$

and $\left[F_1{}^{small}(r_n)\right]_{i,j} = \sum_{k,l} C_{k,i}^{sS^*} C_{l,j}^{pS} N_k^{sS^*} N_l^{pS} (\langle s|\boldsymbol{n}|p\rangle)_z \left(\int_0^{r_n} b_1^{small} r dr\right)$

$$= \sum_{k,l} C_{k,i}^{s\ \ S*} C_{l,j}^{pS} N_k^{sS^*} N_l^{pS} \left(-\frac{8\pi}{\sqrt{3}} \alpha_k^* r_n{}^2\right).$$

(11)

Similarly,

$$[F_2(r_n)]_{i,j} = \left[F_2{}^{large}(r_n)\right]_{i,j} + \left[F_2{}^{small}(r_n)\right]_{i,j}$$

with

$$\left[F_2{}^{large}(r_n)\right]_{i,j}$$

$$= \sum_{k,l} C_{k,i}^{sL^*} C_{l,j}^{pL} N_k^{sL^*} N_l^{pL} (\langle s|\boldsymbol{n}|p\rangle)_z \left(\frac{1}{r_n^3}\int_0^{r_n} b_1^{large} r^4 dr - \int_0^{r_n} b_1^{large} r dr\right)$$

$$= \sum_{k,l} C_{k,i}^{sL^*} C_{l,j}^{pL} N_k^{sL^*} N_l^{pL} \left(-\frac{2\pi}{5\sqrt{3}} r_n{}^2\right)$$

and

$$\left[F_2{}^{small}(r_n)\right]_{i,j}$$

$$= \sum_{k,l} C_{k,i}^{sS^*} C_{l,j}^{pS} N_k^{sS^*} N_l^{pS} (\langle s|\boldsymbol{n}|p\rangle)_z \left(\frac{1}{r_n^3}\int_0^{r_n} b_1^{small} r^4 dr - \int_0^{r_n} b_1^{small} r dr\right)$$

$$= \sum_{k,l} C_{k,i}^{sS^*} C_{l,j}^{pS} N_k^{sS^*} N_l^{pS} \left(\frac{12\pi}{5\sqrt{3}} \alpha_k^* r_n{}^2\right) .$$

(12)

Here, we show the contribution from a single molecular orbital to the electronic term, where $C$ represents the molecular orbital coefficient, $N$ the normalization constant, and $\alpha$ the Gaussian exponential parameter. The superscripts L and S on $C$ and $N$ denote large and small components, respectively, while the subscripts $k$and $l$label atomic orbitals, and $i$ and $j$ label molecular orbitals. We assume uncontracted basis functions, with the summation indices $k$ and $l$ corresponding to the basis function numbers. Using the fact that $\frac{3}{4\pi}b_1(\langle s|\mathbf{n}|p\rangle)_z$ matches the derivative of the electron density at the nucleus, the dominant matrix element of the interaction energy for a one-electron system can be expressed as follows [25-28].

$$\langle s| - e\varphi^{(1)}(\boldsymbol{r})|p\rangle = 4\pi e\boldsymbol{S} \cdot \left(\nabla\psi_s^\dagger\psi_p\right)_{r=0} \tag{13}$$

The Schiff moment **S** is defined as Eq (1).

From this context, the value of $X$, where

$$X = 1/2b_1(\langle s|\boldsymbol{n}|p\rangle)_z, \tag{14}$$

and satisfied with

$$\langle s| - e\varphi^{(1)}(\boldsymbol{r})|p\rangle = 6eXS, \tag{15}$$

has been conventionally used to represent the electronic term.

The approximation in Equation (10) is expected to be accurate for light elements, but for systems containing heavy elements with large nuclei, this approximation may become less reliable. Furthermore, Equation (13) assumes that the nuclear Schiff moment is localized at the center of the nucleus, which undermines its logical consistency [31].

### *2.3.2 Electronic Term of Analytical Representation*

We propose an analytical representation based on Gaussian basis functions centered on the nucleus, which are widely used in quantum chemistry to represent molecular orbitals [30]. In this approach, $U_{sp}(r)$ is naturally described using Gaussian functions. The validity of employing Gaussian basis functions for atomic orbitals is discussed in Section 4.3 by comparing the results with those obtained using a numerical basis set. By adopting this analytical representation, we eliminate the approximation used in Equation (10) of the conventional method. Consequently, the electronic terms are not solely determined by information at the nuclear origin, enabling accurate calculations that are particularly suitable for systems with heavy nuclei and significant NSI. Below, we present an example for electronic state term 2 for the $s_{1/2}$ ($m_j$=+1/2) and $p_{1/2}$ ($m_j$=+1/2) combination, illustrating contributions from both the large and small components. Analytical expressions for the other combinations are provided in appendices and the detailed derivation can be in the supplementary material (SM) [32]. Mathematica [33] was used for this derivation. Following the earlier discussion, we can write

$$[F_2(r_n)]_{i,j} = \left[F_2{}^{large}(r_n)\right]_{i,j} + \left[F_2{}^{small}(r_n)\right]_{i,j}$$

with

$$\left[F_2{}^{large}(r_n)\right]_{i,j} = \sum_{k,l} C_{k,i}^{s\ \ L*} C_{l,j}^{pL} N_k^{sL^*} N_l^{pL} \left( -\frac{3e^{-\alpha'_{k,l}r_n{}^2} r_n + 2\alpha'_{k,l} r_n{}^3}{4\alpha'_{k,l}{}^2 r_n{}^3} - \frac{3\sqrt{\pi} erf\left[\sqrt{\alpha'_{k,l}} r_n\right]}{8\alpha'_{k,l}{}^{5/2} r_n{}^3} \right) \frac{4\pi}{3\sqrt{3}}$$

and

$$\left[F_2{}^{small}(r_n)\right]_{i,j} = \sum_{k,l} C_{k,i}^{s\ \ S*} C_{l,j}^{pS} N_k^{sS^*} N_l^{pS} \left( \frac{\alpha_k\left(3\alpha'_{k,l} - 2\beta\right)}{\alpha'_{k,l}{}^2} - \frac{3\alpha_k e^{-\alpha'_{k,l}r_n{}^2}\left(5\beta_l + \alpha'_{k,l}(-3 + 2\beta_l r_n{}^2)\right)}{2\alpha'_{k,l}{}^3 r_n{}^2} + \frac{3\alpha_k(-3\alpha'_{k,l} + 5\beta_l)\sqrt{\pi} erf\left[\sqrt{\alpha'_{k,l}} r_n\right]}{4\alpha'_{k,l}{}^{7/2} r_n{}^3} \right) \frac{4\pi}{3\sqrt{3}}. \quad (16)$$

Here, $\alpha_k$ and $\beta_l$ denote the exponential parameters of the bra and ket basis functions, respectively; $a'_{k,l}$ is their sum, and $erf[x]$ represents the error function. By expanding $U_{sp}(r)$—expressed using Gaussian basis functions—in a power series and retaining only the lowest-order term, we obtain an expression equivalent to the conventional formulation described above, which are provided in appendices. For both the conventional and analytical representations, the electronic terms for the many-electron contributions are calculated as

$$F_{1,HF}(r_n) = \sum_{i=1}^{N_e}[F_1(r_n)]_{i,i}, \quad F_{2,HF}(r_n) = \sum_{i=1}^{N_e}[F_2(r_n)]_{i,i} \quad (17)$$

at the Hartree-Fock (HF) level, and

$$F_{1,CCSD}(r_n) = \sum_{i=1}^{N_{MO}}\sum_{j=1}^{N_{MO}}[F_1(r_n)]_{i,j}\gamma_{j,i}\,, \; F_{2,CCSD}(r_n) = \sum_{i=1}^{N_{MO}}\sum_{j=1}^{N_{MO}}[F_2(r_n)]_{i,j}\gamma_{j,i} \; (18)$$

at the CCSD level. Here, $\gamma_{j,i}$ denotes the first-order density matrix obtained from the CCSD calculation, and $N_e$ and $N_{\mathrm{MO}}$ denote the number of electrons and molecular orbitals, respectively. By substituting the $F_1$ and $F_2$ functions obtained in this study into the left-hand side of Eq. (7) and performing the integration over the nuclear coordinates using the CP-violating proton density, the interaction energy observed in experiments can be evaluated.

*2.3.3 Electronic Terms in Alternative Representations*

Several methods have been proposed to account for the effects of the electronic state in NSI without relying solely on nuclear-centered information [25,34]. In many of these approaches, the electronic parameters are reformulated so that the definition of the NSM remains unchanged, and the interaction energy is given as the product of the NSM and a scalar electronic parameter. However, in Eq. (4) of Ref. [25] approximations regarding the nuclear structure are introduced in the derivation, yielding an empirical formula that does not correspond to our Eq. (7). In contrast, Eq. (16) in Ref. [25] starts from the same interaction Hamiltonian as our Eq. (7) and introduces the concept of local dipole moment (LDM) to account for the influence of higher-order moments beyond the conventional NSM term [25]. Nevertheless, the electronic parameters representing these higher-order terms are derived from analytical solutions for hydrogen-like atoms [25], and it remains unclear whether they are applicable to molecular systems.

In addition, when expressing the NSI energy, using the NSM definition from Eq. (4) for the nuclear part is based on approximations that may break down in certain cases. For example, if the electronic term—as shown in this study—exhibits a curve that cannot be described by a quadratic function of $r_n$ (as in the case of LrF), the NSM representation would fail. Even when one insists on the NSM representation to conveniently separate nuclear and electronic contributions, fitting the analytical electronic term to a quadratic function in the nuclear region appears to be the most accurate approach. However, with current computational techniques, it is possible to numerically integrate the electronic function together with the nuclear wavefunction, bypassing the NSM representation altogether and yielding a more accurate interaction energy. Therefore, in this study, we express the electronic term as a function of the nuclear coordinate rather than as a scalar quantity. The advantage of our approach is that it is more straightforward and accurate than other methods due to the absence of a Maclaurin expansion, and we consider it a benchmark relative to previously proposed approaches.

## 3. Computational Methods

In this study, we performed benchmark calculations on the $^{205}$TlF [35], $^{225}$RaO [36], and $^{256}$LrF molecules. $^{205}$TlF is attractive due to its chemical stability, high polarity and polarizability, simple electronic structure, enhanced interaction from the $^{205}$Tl nucleus, and straightforward nuclear structure (a single unpaired proton in $^{205}$Tl or a proton hole in $^{19}$F) [37]. These features have also made it a popular choice in experimental studies. $^{225}$RaO is expected to exhibit NSI larger than that predicted by the conventional Z-scaling approximation and is purported to have an interaction energy 500 times greater than that of $^{205}$TlF [36]. Although $^{256}$Lr is a superheavy element and forming a large number of its molecules is challenging, its large atomic number suggests that $^{256}$LrF may exhibit a significant NSI effect.

In addition to Dirac-Hartree-Fock (DHF) calculations, we performed four-component relativistic CCSD calculations to compute the electronic terms. For this purpose, one-particle density matrices with the λ-CCSD method were utilized. The electronic calculations were carried out using UTChem [30,38-40] and DIRAC21.1 [41,42]. DHF calculations employing the four-component Dirac–Coulomb Hamiltonian were performed with UTChem, while CCSD calculations were conducted with RELCCSD program [43,44] implemented in DIRAC. In the CCSD calculations, all electrons were correlated, and 604, 602, and 640 virtual spinors were included in the correlation treatment for TlF, RaO, and LrF, respectively. For each molecule, we employed a finite nuclear model with a Gaussian charge distribution, using nuclear radii determined from an empirical formula that estimates the charge radius from the mass number [45]. The empirical formula for the root-mean-squared (RMS) charge radius $\langle r_n^2\rangle^{1/2}$ and the relationship between the nuclear radius $R$ and $\langle r_n^2\rangle^{1/2}$ are given by

$$\langle r_n^2\rangle^{1/2} = 0.836A^{1/3} + 0.570 \quad \text{fm} \tag{19}$$

and

$$R = \sqrt{\frac{5}{3}} \times \langle r_n^2\rangle^{1/2} \, . \tag{20}$$

The corresponding numerical values of the above factors for different atoms of our interest are given in Table 1.

**Table 1.** Root-mean-square nuclear radii $\boldsymbol{\langle r_n^2\rangle^{1/2}}$ and nuclear radii $\boldsymbol{R}$ for $^{205}$**Tl**, $^{225}$**Ra**, and $^{255}$**Lr** atoms.

| | $\langle r_n^2\rangle^{1/2}$ (fm) | $R$ (fm) | $R$ (a.u.) |
|---|---|---|---|
| $^{205}$Tl | 5.4994 | 7.0994 | $1.3416\times10^{-4}$ |
| $^{225}$Ra | 5.6547 | 7.3000 | $1.3795\times10^{-4}$ |
| $^{255}$Lr | 5.8783 | 7.5889 | $1.4341\times10^{-4}$ |

In addition, the equilibrium internuclear distances—2.0844 Å for TlF and 2.038 Å for RaO—were taken from experimental measurements [46] and theoretical calculations [47], respectively. For LrF, a theoretically estimated value of 2.005092 Å, obtained at the B3LYP/Dyall.v2z level of theory, was used.

Electronic term calculations are highly dependent on the chosen basis set, making the selection of basis functions crucial [37]. For heavy atoms, we employed basis sets composed primarily of even-tempered (ET) functions, with polarization functions added from the Dyall.cv4z set [48-51]. The ET basis functions determine the exponential parameters via a geometric progression, which allows a flexible description of the near-nucleus region when these parameters are appropriately chosen. The ET basis sets of heavy atoms (Tl, Ra, Lr) were tuned using Quiney et al.'s validation method [37] to accurately reproduce the numerical radial functions in the nuclear region. For light atoms (O, F), we used the Dyall.cv4z basis set. Details regarding basis sets are provided in SM [52]. The numerical radial functions were obtained using a modified version of GRASP2K [53]. To investigate basis-set dependence, we also performed calculations using the Dyall.cv4z basis sets for heavy elements [48-51] and compared these results with those obtained using the ET basis set.

## 4. Results and Discussion

### *4.1. Results of conventional numerical approach (Values of X)*

Before discussing the electronic terms, we first examine the consistency of the conventional representation. Table 2 compares the values of $X$ obtained using Eq. (14) with those reported in the previous study (Ref. 32), under the computational conditions employed in this work. Although the values of $X$ obtained in our calculation for RaO are negative, the absolute values $|X|$ are presented in the table to facilitate comparison with the previous results.

In the previous study (Ref. 32), the finite nuclear size effect was treated using a correction formula derived from a Maclaurin expansion of the analytical solution for a hydrogen-like atom. As a result, correction factors of 1.13, 1.16, and 1.16—denoted as $\bar{r}_{sp}$ in Ref. 32—were introduced for TlF, RaO, and AcF, respectively. The corrected values are obtained by dividing by these factors and are listed in Table 2 as *Corrected* $|X|$.

Since the ET basis set used in the present work was constructed with reference to the basis set employed in Ref. 37 for TlF, the uncorrected $|X|$ values agree well with our calculated results. In contrast, when the Dyall basis set is used, a larger discrepancy is observed. The basis set used in the previous study for RaO was designed to reproduce experimental excitation energies and therefore may not be well suited for describing the electronic structure within the nuclear region. Consequently, this result is closer to the value obtained in our calculations with the Dyall basis set. As no previous results are available for LrF, the data for AcF are included for reference; these values are found to be close to our result obtained with the ET basis set.

The correction factors introduced in Ref. 32 correspond to differences of approximately 13–16%, which are considerably smaller than the differences between the conventional and analytical representations obtained with the ET basis set in the present work (50% for RaO and 300% for LrF). The origin of this discrepancy will be discussed in detail later, but it is likely attributable to the fact that the previous studies relied on approximations based on Maclaurin expansions truncated at low order.

**Table 2. Comparison of the conventional numerical representation *X* with the previous results reported in Ref. 32.**

| Molecule | $\|X\|$ | Corrected $\|X\|$ | Ref, |
|---|---|---|---|
| TlF | 8747 | 7741 | [32][a] [37] |
| TlF | 8928 | - | This work (HF/ET) |
| TlF | 6306 | - | This work (HF/Dyall) |
| RaO | 7532 | 6521 | [32][a] [35] [47] |
| RaO | 15492 | - | This work (HF/ET) |
| RaO | 8731 | - | This work (HF/Dyall) |
| AcF | 26667 | 23333 | [32][a] [54] |
| LrF | 31517 | - | This work (HF/ET) |
| LrF | 5605 | - | This work (HF/Dyall) |

[a] In Ref. 32, $W_s = 6X$ was reported instead of $X$.

### *4.2. Numerical representation (conventional) vs. Analytical representation*

Below, we present the plots of electronic term 1 and electronic term 2 as Figure 1 and 2, obtained using both the conventional and analytical representations at the HF and CCSD levels with the ET basis set. We restrict the plots to the nuclear region ($0.0 \leq r_n \leq 1.5\times10^{-4}$ a.u.) because the interaction energy is calculated from the product of the electronic term and the proton density $\rho(\mathbf{r_n})$, which approaches zero around $1.5\times10^{-4}$ a.u.. Although contributions from the entire $r_n$ range—from the nuclear center to the surface—are important, the conventional representation is a lowest-order Maclaurin expansion of the analytical representation, so they always agree well with each other near the nuclear origin. For the calculation of the interaction energy, the electronic terms are integrated with the weight $\rho(r_n)r_n^3$ (where $r_n^2$ arises from the Jacobian and $r_n$ originates from the original formula in Eq. (7)), indicating that contributions from larger $r_n$ are more significant. Therefore, we evaluate the approximation accuracy of the conventional representation by comparing it with the analytical representation at the nuclear radius. The deviation is calculated as the relative difference from the analytical value, which is taken as 100%. Table 3 presents the electronic term values at the nuclear radius $R$ for each molecule, along with the difference ratios between the conventional and analytical representations. It should be noted that only in RaO the sign of the electronic term is reversed compared with the other molecules. This is likely due to the difference in whether the s and p orbitals mix in phase or out of phase in the molecular orbitals. When discussing the interaction energy, the sign itself becomes irrelevant because the absolute value of the interaction energy is considered. However, since the sign may be related to whether the orbital is bonding or antibonding, we retain the sign obtained from the calculations and present the results accordingly in the tables and figures.

**Table 3.** Values of electronic term 1 and 2 at the nuclear radius for $^{205}$TlF, $^{225}$RaO, and $^{256}$LrF, along with the difference ratios between the conventional and analytical representations.

| | HF Analytical (a.u.) | HF Conventional (a.u.) | HF Difference ratio* (%) | CCSD Analytical (a.u.) | CCSD Conventional (a.u.) | CCSD Difference ratio* (%) |
|---|---|---|---|---|---|---|
| $^{205}$TlF, $F_1(r_n)$ | $1.43\times10^{-4}$ | $1.61\times10^{-4}$ | 12.2 | $1.13\times10^{-4}$ | $1.26\times10^{-4}$ | 12.1 |
| $^{225}$RaO, $F_1(r_n)$ | $-1.91\times10^{-4}$ | $-2.95\times10^{-4}$ | 54.1 | $-1.63\times10^{-4}$ | $-2.63\times10^{-4}$ | 61.6 |
| $^{256}$LrF, $F_1(r_n)$ | $1.51\times10^{-4}$ | $6.48\times10^{-4}$ | 328.5 | $1.66\times10^{-4}$ | $6.50\times10^{-4}$ | 290.8 |
| $^{205}$TlF, $F_2(r_n)$ | $-8.81\times10^{-5}$ | $-9.64\times10^{-5}$ | 9.3 | $-6.93\times10^{-5}$ | $-7.57\times10^{-5}$ | 9.3 |
| $^{225}$RaO, $F_2(r_n)$ | $1.19\times10^{-4}$ | $1.77\times10^{-4}$ | 48.8 | $1.01\times10^{-4}$ | $1.58\times10^{-4}$ | 55.9 |
| $^{256}$LrF, $F_2(r_n)$ | $-1.05\times10^{-4}$ | $-3.89\times10^{-4}$ | 269.6 | $-1.08\times10^{-4}$ | $-4.02\times10^{-4}$ | 271.3 |

* The difference ratio, expressed as a percentage, is calculated as $\left(\left|\frac{\text{Conventional value}}{\text{Analytical value}}\right| - 1\right) \times 100.$

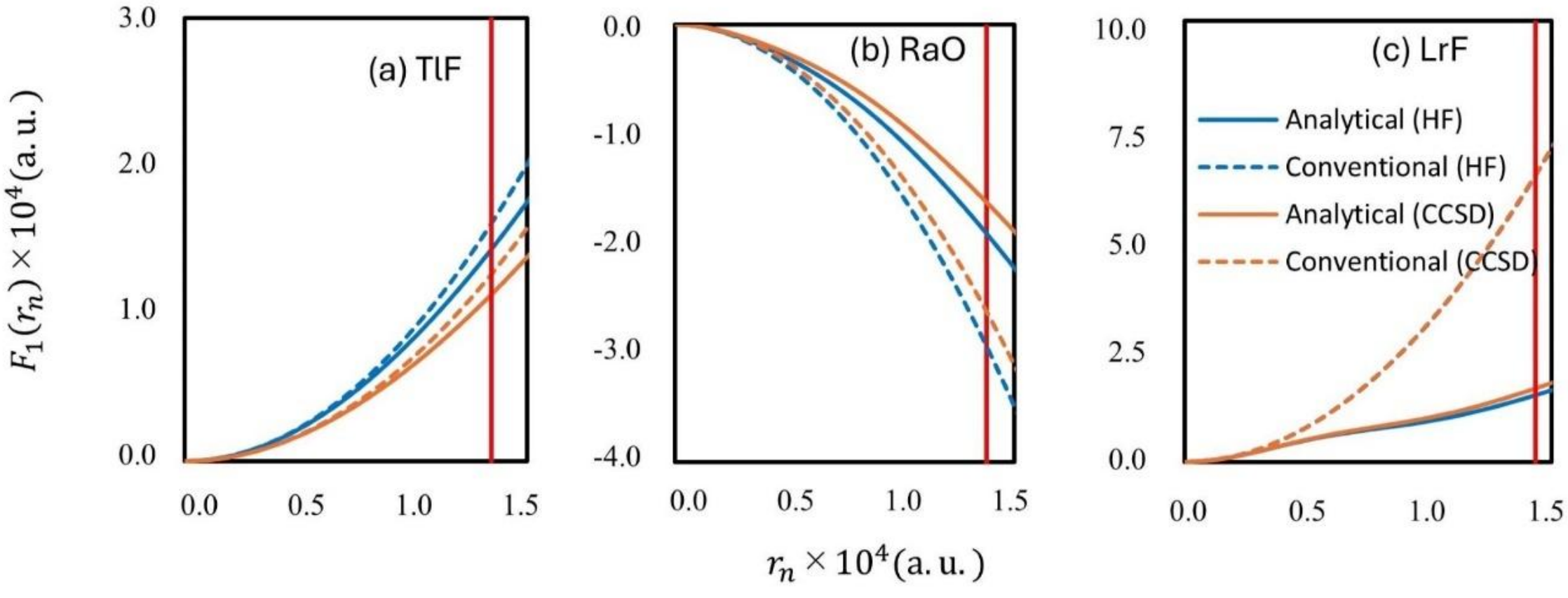

**Figure 1.** Plots of electronic term 1 for (a) $^{205}$TlF, (b) $^{225}$RaO, and (c) $^{256}$LrF. The red vertical line indicates the nuclear radius $R$ of the $^{205}$Tl, $^{225}$Ra, and $^{256}$Lr atom.

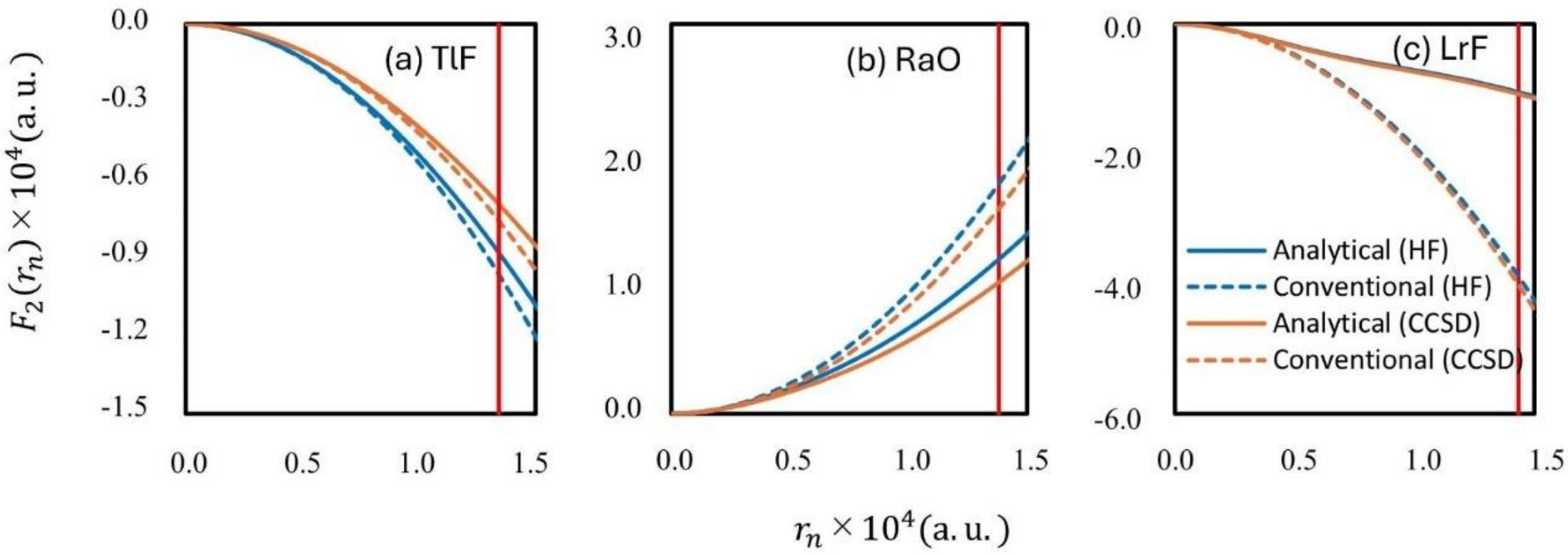

**Figure 2.** Plots of electronic term 2 for (a) $^{205}$TlF, (b) $^{225}$RaO, and (c) $^{256}$LrF. The red vertical line indicates the nuclear radius $R$ of the $^{205}$Tl, $^{225}$Ra, and $^{256}$Lr atom.

Overall, the results show that the conventional representation consistently overestimates the electronic term values compared to the analytical representation for all molecules and for both electronic terms. For $^{205}$TlF, the conventional representation overestimates electronic term 1 by 12% and electronic term 2 by 9%, for both HF and CCSD levels, indicating relatively good accuracy for this molecule. In contrast, for $^{225}$RaO, the overestimation is much larger: electronic term 1 is overestimated by 54% (HF) and 62% (CCSD), while electronic term 2 is overestimated by 49% (HF) and 56% (CCSD). This suggests that using the analytical representation significantly improves calculation accuracy for $^{225}$RaO. For $^{256}$LrF, the overestimation is even more pronounced, with electronic term 1 overestimated by 328% (HF) and 291% (CCSD), and electronic term 2 by 270% (HF) and 271% (CCSD). These findings clearly demonstrate that the analytical representation is especially important for achieving accurate calculations in $^{256}$LrF.

Figure 3 compares the three molecules at the HF level using the absolute magnitudes of the electronic terms. Near the nuclear center, both the conventional and analytical representations yield the ordering $^{205}$TlF < $^{225}$RaO < $^{256}$LrF. This is consistent with the Z-scaling argument, which predicts that the atomic EDM increases with the cube of the atomic number Z. However, because Z-scaling applies strictly to atoms and does not account for contributions from light atoms in these molecules, this ordering may not be entirely accurate. As $r_n$ increases, the analytical representation shows a change in the ordering of the electronic terms—from $^{205}$TlF < $^{256}$LrF < $^{225}$RaO to $^{256}$LrF < $^{205}$TlF < $^{225}$RaO. This shift, which is observed only with the analytical representation, is attributed to distortions in the electronic term of $^{256}$LrF. This distortion arises from the superposition of the large- and small-component functions with different shapes and therefore does not indicate numerical instability. In contrast, the conventional representation, being always quadratic, maintains a consistent ordering across the entire $r_n$ range. Thus, the analytical representation reveals behavior in the electronic terms that the conventional method cannot predict. Furthermore, these results suggest that $^{256}$LrF may be less suitable for EDM observation than previously expected, underscoring the utility of the analytical representation in identifying promising experimental systems for EDM measurements.

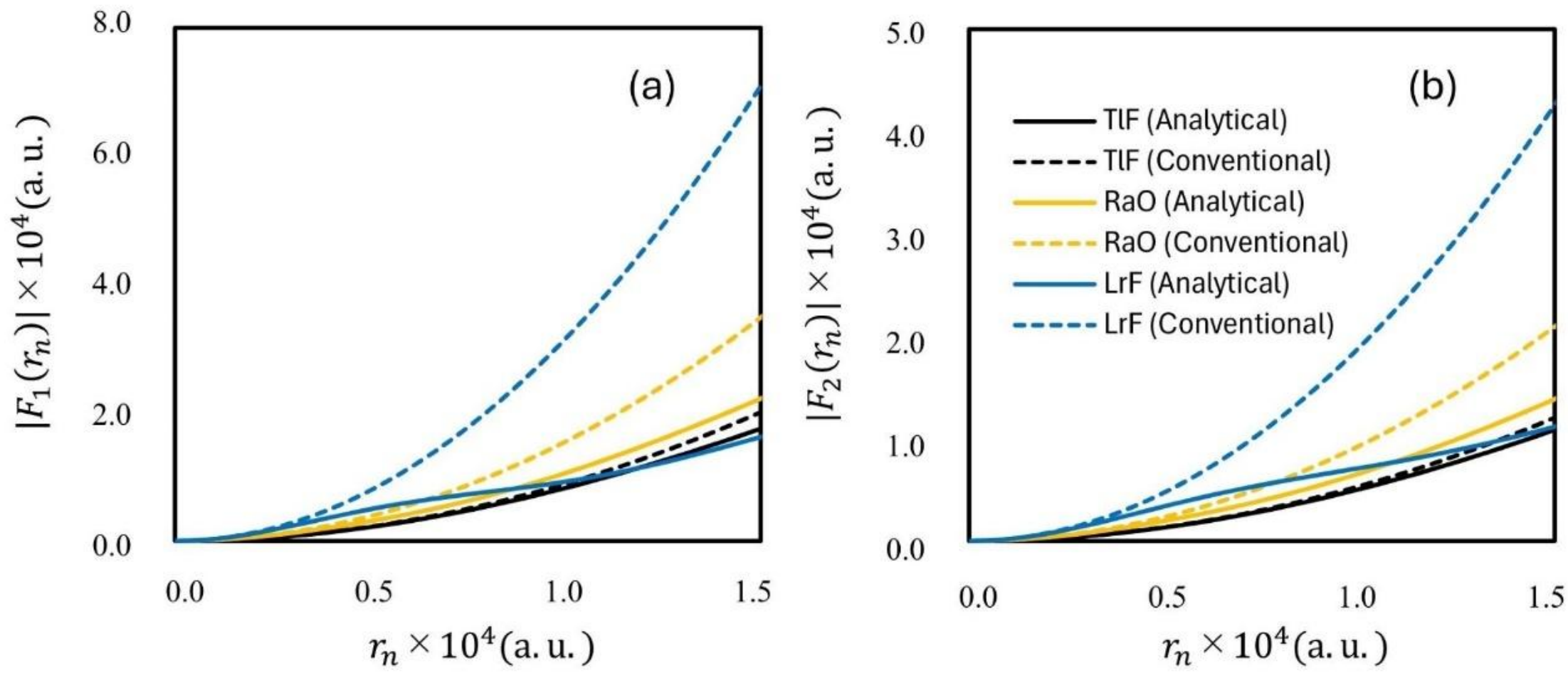

**Figure 3.** Comparison of electronic (a) term 1 and (b) term 2 for $^{205}$TlF, $^{225}$RaO, and $^{256}$LrF.

### *4.3 Higher-Order Maclaurin Expansions*

In the nuclear-coordinate integration required to obtain the final interaction energy, expressing the electronic term as a polynomial expansion in $r_n$may be computationally convenient, because analytical integration can then be applied. Therefore, to examine whether higher-order Maclaurin expansions can serve as effective approximations for the electronic term, we performed calculations for the RaO molecule using fourth- and tenth-order Maclaurin expansions in addition to the conventional second-order term (Fig. 4).

As shown in Fig. 4, higher-order expansions reproduce the analytical representation over a wider range of $r_n$. However, larger errors appear in the nuclear-radius region (around $r_n \approx 1.3 \times 10^{-4}$ a.u.) compared with the second-order case. This indicates that the nuclear radius is too large for Maclaurin expansions up to about the tenth order to accurately approximate the analytical expression.

This observation also raises concerns about correction methods used in previous studies that rely on low-order Maclaurin expansions of the point-nucleus approximation (Table 2). However, the correction factor introduced in Ref. 32 is not derived as an approximation to Eq. (7), which forms the basis of our analytical representation. Instead, it is based on a different empirical formula, and therefore a direct comparison is not entirely appropriate.

Nevertheless, the present results indicate that, assuming Eq. (7) to be valid, Maclaurin expansions of moderate order cannot accurately reproduce the analytical solution up to the nuclear radius. Consequently, the analytical representation without any truncation of a Maclaurin expansion should be considered more reliable.

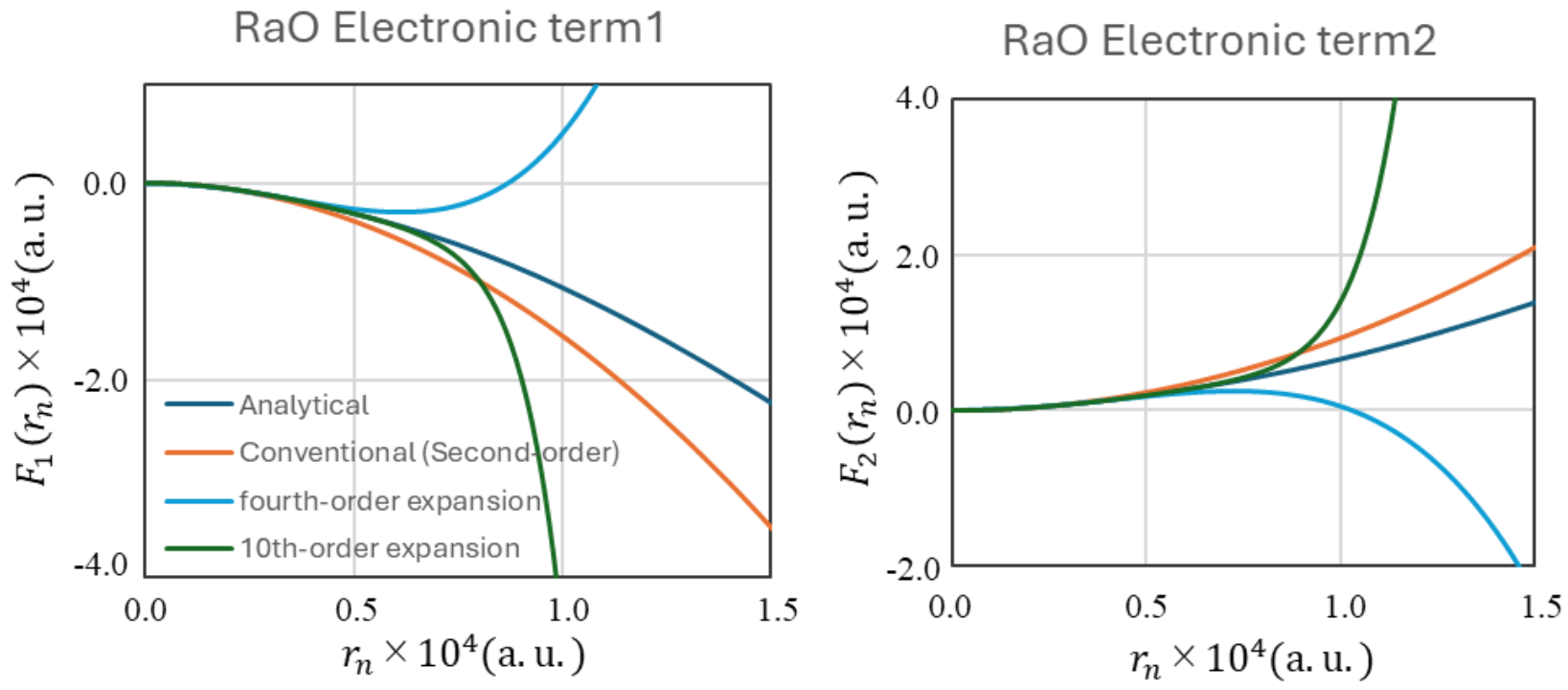

**Figure 4.** Comparison of the analytical representation with second-(conventional), fourth-, and tenth-order Maclaurin expansions for the electronic terms of RaO.

### *4.4 Evaluation of Basis-Set Dependence*

To assess the basis-set dependence in the analytical representation, we compare the electronic term calculations performed using ET basis functions and the fully Dyall.cv4z basis set. Since Section 4.1 shows that electronic term 1 and electronic term 2 exhibit similar trends, we use electronic term 1 for this evaluation. Figure 4 presents the plots of electronic term 1 for $^{205}$TlF, $^{225}$RaO, and $^{256}$LrF, respectively, for both the ET and Dyall.cv4z basis sets. Table 4 summarizes the values of the electronic term 1 at the nuclear radius for each molecule and the percentage differences between the two basis sets. Table 4 shows that for $^{205}$TlF, the conventional representation exhibits a basis-set variation of 41%, while the analytical representation reduces this variation to 16%. Similar trends are observed for the other molecules: for $^{225}$RaO, the variation decreases from 78% to 13%, and for $^{256}$LrF, it drops dramatically from 462% to 51%. These results indicate that the analytical representation is less sensitive to the choice of basis set than the conventional representation. Consequently, using the analytical representation allows for more precise and reliable analyses that are less dependent on the specific basis functions employed.

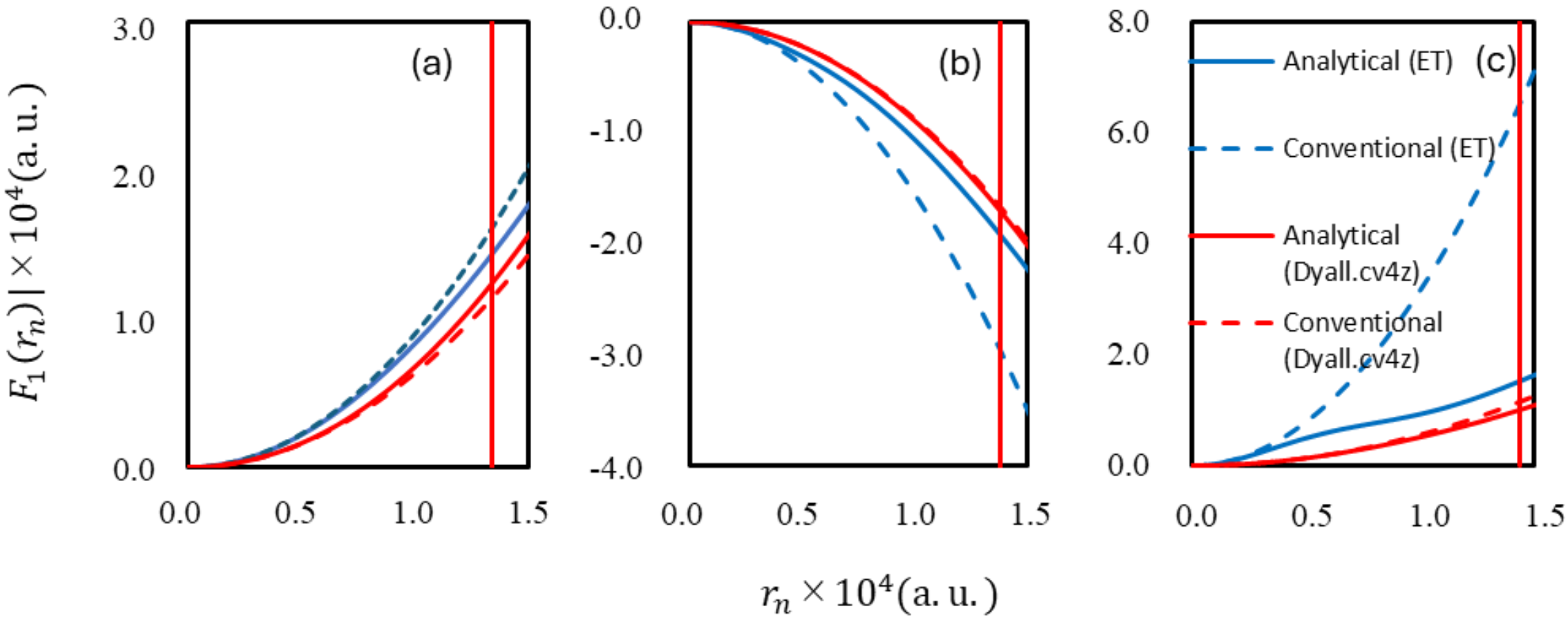

**Figure 5.** Comparison of ET basis sets and Dyall.cv4z basis sets in the electronic term 1 for (a) $^{205}$TlF, (b) $^{225}$RaO, and (c) $^{256}$LrF.

**Table 4**. Values of electronic term 1 at the nuclear radius for each molecule using different basis sets, along with the percentage differences observed between them.

| | Analytical ET (a.u.) | Analytical Dyall.cv4z (a.u.) | Analytical Difference ratio* (%) | Conventional ET (a.u.) | Conventional Dyall.cv4z (a.u.) | Conventional Difference ratio* (%) |
|---|---|---|---|---|---|---|
| $^{205}$TlF | $1.43\times10^{-4}$ | $1.24\times10^{-4}$ | 15.6 | $1.61\times10^{-4}$ | $1.14\times10^{-4}$ | 41.5 |
| $^{225}$RaO | $-1.91\times10^{-4}$ | $-1.70\times10^{-4}$ | 12.5 | $-2.95\times10^{-4}$ | $-1.66\times10^{-4}$ | 77.7 |
| $^{256}$LrF | $1.51\times10^{-4}$ | $1.00\times10^{-4}$ | 50.8 | $6.48\times10^{-4}$ | $1.15\times10^{-4}$ | 462.3 |

* The difference ratio, expressed as a percentage, is calculated as $\left(\left|\frac{\text{ET value}}{\text{Dyall.cv4z value}}\right| - 1\right) \times 100.$

### *4.5 High-Precision Electronic Term Calculations via Improved Basis Functions*

In the former Section, we demonstrated that the analytical representation reduces basis-set dependence compared to the conventional representation, highlighting its advantages. However, for the $^{256}$LrF molecule, even the analytical representation exhibits a basis-set-induced variation of 51%. Here, we aim to improve the accuracy of electronic term calculations by refining the basis set functions. Additionally, we evaluate the accuracy of the ET basis functions by comparing the results obtained with the improved basis functions. Due to computational cost constraints, $^{225}$RaO was chosen as this benchmark molecule.

To analyze the differences in electronic term calculations arising from the choice of basis functions, we compared the ET and Dyall.cv4z basis functions against numerical basis functions. We selected the $6p_{1/2}$ orbital of the $^{225}$Ra atom, which is the primary component of the molecular orbital that shows the maximum contribution in the electronic term calculation for $^{225}$RaO. Although the nuclear region is usually the most important for electronic term calculations, we compared the radial functions both in the nuclear region and across a wider range that includes the valence region for a more detailed analysis. Since a direct comparison did not reveal significant differences (see SM [32]), indicating that Gaussian basis functions are sufficiently accurate, we plotted the differences between the radial functions obtained using the numerical basis functions and those obtained with each basis set.

For the $6p_{1/2}$ orbital of $^{225}$Ra, Figures 5 shows the differences in the radial functions within the nuclear region, while Figures 6 presents the comparisons over a wider range including the valence region.

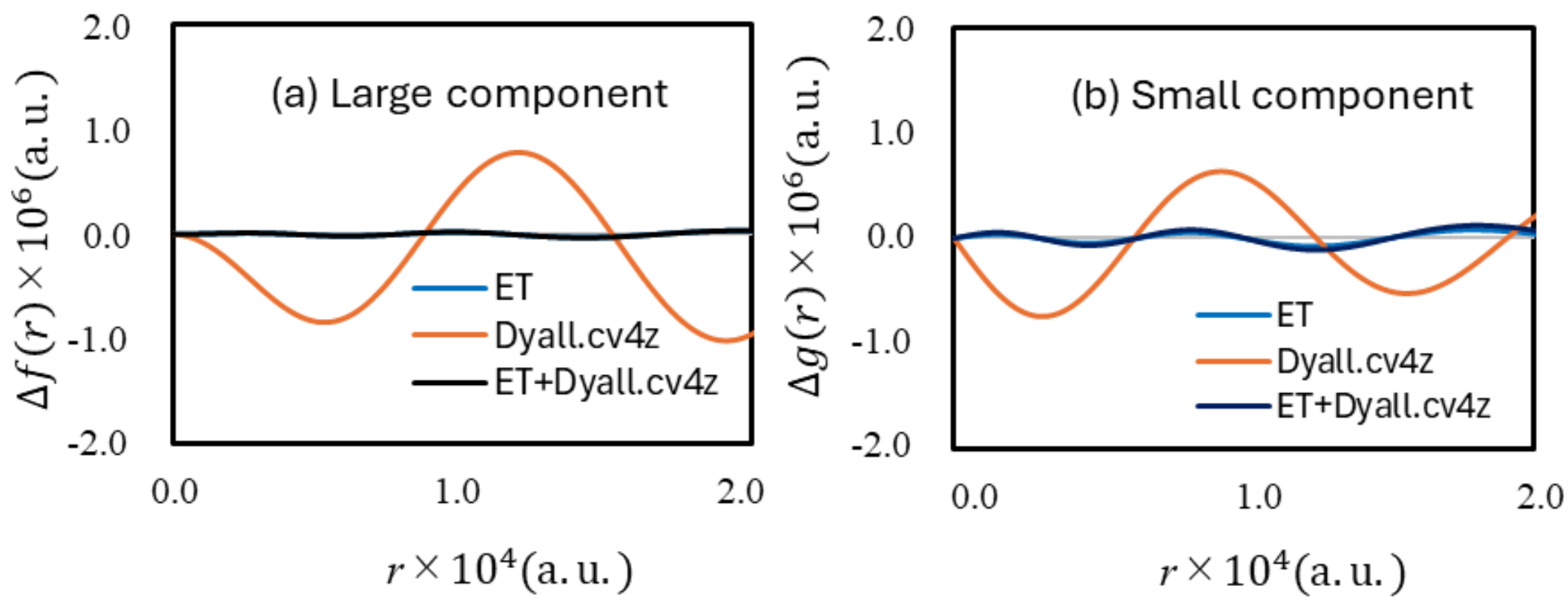

**Figure 6**. Difference plot of the radial function for the (a) large and (b) small component of the $^{225}$Ra $6p_{1/2}$ orbital in the nuclear region, comparing each basis set with the numerical basis function.

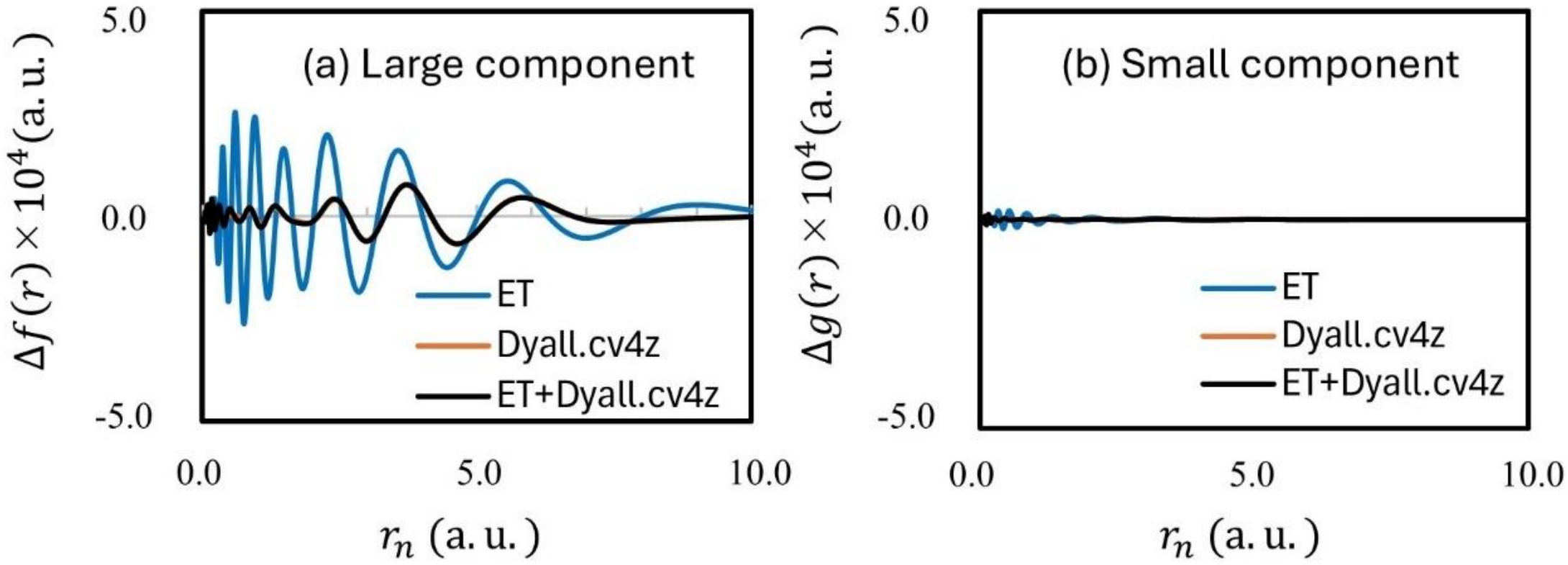

**Figure 7**. Difference plot of the radial function for the (a) large and (b) small component of the $^{225}$Ra $6p_{1/2}$ orbital over a wider range including the valence region, comparing each basis set with the numerical basis function.

Figures 6 and 7 show that ET basis functions accurately describe the radial functions in the nuclear region, while Dyall.cv4z performs better in the valence region. This suggests that Quiney et al.'s tuning [37] is effective in the nuclear region but less so in the valence region. Because diatomic molecules require accurate descriptions in both regions to properly capture chemical bonding and its effects, the ideal basis set must perform well overall. Therefore, we developed a new basis set (ET+Dyall.cv4z) by selecting the exponential parameters that strongly contribute to the description of the atomic orbitals in the nuclear and valence regions from the ET and Dyall.cv4z basis sets, respectively. As shown in Figures 6 and 7, the new ET+Dyall.cv4z basis set agrees with the ET basis in the nuclear region and with Dyall.cv4z in the valence region, demonstrating its satisfactory performance.

Electronic term 1 and electronic term 2 were plotted using the ET+Dyall.cv4z basis set and compared with the results from the ET and Dyall.cv4z basis sets in Figure 8. Figures 8 (a) and (b) show electronic term 1 and electronic term 2, respectively, for both the conventional and analytical representations across these basis sets. For both electronic terms, the analytical representation using the ET+Dyall.cv4z basis set closely matches the results obtained with the ET basis set. This indicates that ET basis functions, which provide a more flexible description of the nuclear region, are well-suited for analytical representation calculations—especially compared to Dyall.cv4z, which performs better in the valence region. Therefore, Quiney et al.'s tuning [37] is highly effective, and using ET basis functions is a very reasonable approach in the analytical representation.

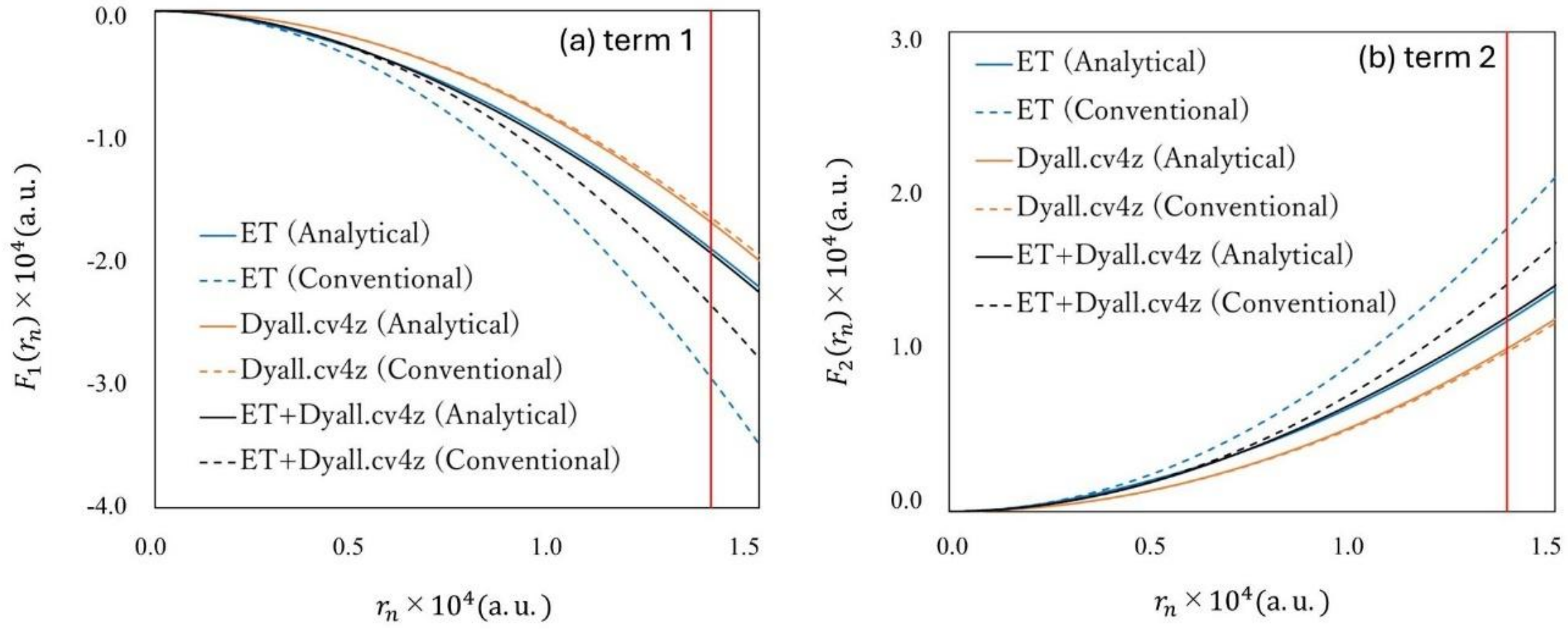

**Figure 8**. Plots of electronic term 1 (a) and electronic term 2 (b) for the $^{225}$RaO molecule using each basis set.

### *4.6 Impact on the Interaction Energy*

Thus far, we have compared the conventional and analytical approaches using the values of the electronic terms at the nuclear radius, $F_1(R)$and $F_2(R)$, as a simplified indicator of the interaction energy. However, the quantity of primary interest is the resulting interaction energy $E$ itself.

A rigorous evaluation of the interaction energy requires the determination of the nuclear wavefunction using an appropriate nuclear-structure model, such as the nuclear shell model, the harmonic oscillator model, or more advanced approaches. Since the present work focuses on electronic-structure calculations, such nuclear calculations are beyond the scope of this study. To facilitate future theoretical investigations by researchers specializing in nuclear wavefunction calculations, the functions $F_1$ and $F_2$ are provided numerically in the Supplementary Information (SI).

Here, without performing explicit nuclear wavefunction calculations, we provide a rough estimate of the impact on the interaction energy based on the characteristics of the obtained $F_1$ and $F_2$ functions. The ratios between the conventional and analytical representations of both $F_1$ and $F_2$ near the nuclear radius are found to be similar. Therefore, even when performing the integration corresponding to their differences, the resulting interaction energies are expected to scale by approximately the same factor between the two approaches.

Furthermore, according to Eq. (22), the interaction energy is obtained by integrating the electronic terms multiplied by $r_n^3$ (where one power of $r_n$ originates from the original Hamiltonian and $r_n^2$ from the Jacobian) and the nuclear charge density. This suggests that contributions from regions at larger $r_n$ may play an important role. However, the quantitative importance of these contributions can only be assessed through explicit calculations. Following Ref. 32, we therefore assume that the interaction energy depends on an average of the contributions at the origin and at the nuclear radius. Since the conventional and analytical representations are identical at the origin, no error arises there.

Under this assumption, the resulting error is approximately half of that at the nuclear radius, leading to an overestimation of about 28% for RaO and 135% for LrF when using the conventional representation. Under this estimate, the error for RaO approaches the 15% value reported in Ref. 32, while it remains substantially larger for LrF.

## Conclusion

In this study, we introduced an analytical representation of the electronic terms in the NSI using Gaussian basis functions. Benchmark calculations on $^{205}$TlF, $^{225}$RaO, and $^{256}$LrF revealed significant differences at the nuclear radius between the conventional and analytical representations, indicating that the conventional approach lacks sufficient accuracy for heavy-atom systems and underscoring the importance of the analytical method. Even with higher-order Maclaurin expansions (e.g., up to the 10th order), the analytical representation cannot be accurately reproduced in the nuclear-radius region. This suggests that a Maclaurin expansion, which relies on local information at the origin, is not suitable for describing the electronic term over the finite nuclear region.

We further evaluated the basis-set dependence by comparing results obtained with ET basis functions and Dyall.cv4z basis functions. The analytical representation exhibited much lower sensitivity to the choice of basis set than the conventional approach. Finally, we developed a new basis set (ET+Dyall.cv4z) that achieves high accuracy in both the nuclear and valence regions, demonstrating that ET basis functions yield high accuracy in electronic term calculations when appropriately produced.

In summary, our results demonstrate that the analytical representation is more effective than the conventional approach for representing electronic terms, particularly for high-precision calculations in molecular systems containing heavy atoms. The numerical data for these analytical electronic terms, summarized in SM [55], can be used to accurately obtain the interaction energy when integrated with nuclear wavefunction information.

## Appendixes

### *1. Electronic terms of conventional representation*

(1) The cases for the combinations $\left(s_{1/2}^{+1/2}, p_{1/2}^{+1/2}\right)$, $\left(s_{1/2}^{-1/2}, p_{1/2}^{-1/2}\right)$, $\left(p_{1/2}^{+1/2}, s_{1/2}^{+1/2}\right)$, and $\left(p_{1/2}^{-1/2}, s_{1/2}^{-1/2}\right)$. For the ($p$, $s$) combinations, the expressions can be obtained by interchanging the roles of the orbitals in the corresponding ($s$, $p$) cases.

$$[F_1(r_n)]_{i,j} = \left[F_1{}^{large}(r_n)\right]_{i,j} + \left[F_1{}^{small}(r_n)\right]_{i,j} \tag{22}$$

$$\left[F_1{}^{large}(r_n)\right]_{i,j} = \sum_{k,l} C_{k,i}^{sL^*} C_{l,j}^{pL} N_k^{sL^*} N_l^{pL} (\langle s|\boldsymbol{n}|p\rangle)_z \int_0^{r_n} b_1^{large} r dr$$

$$= \sum_{k,l} C_{k,i}^{sL^*} C_{l,j}^{pL} N_k^{sL^*} N_l^{pL} \frac{2\pi}{3\sqrt{3}} r_n{}^2 \tag{23}$$

$$\left[F_1{}^{small}(r_n)\right]_{i,j} = \sum_{k,l} C_{k,i}^{sS^*} C_{l,j}^{pS} N_k^{sS^*} N_l^{pS} (\langle s|\boldsymbol{n}|p\rangle)_z \left(\int_0^{r_n} b_1^{small} r dr\right)$$

$$= \sum_{k,l} C_{k,i}^{sS^*} C_{l,j}^{pS} N_k^{sS^*} N_l^{pS} \left(-\frac{8\pi}{\sqrt{3}} \alpha_k^* r_n{}^2\right) \tag{24}$$

$$[F_2(r_n)]_{i,j} = \left[F_2{}^{large}(r_n)\right]_{i,j} + \left[F_2{}^{small}(r_n)\right]_{i,j} \tag{25}$$

$$\left[F_2{}^{large}(r_n)\right]_{i,j} = \sum_{k,l} C_{k,i}^{sL^*} C_{l,j}^{pL} N_k^{sL^*} N_l^{pL} (\langle s|\boldsymbol{n}|p\rangle)_z \left(\frac{1}{r_n^3}\int_0^{r_n} b_1^{large} r^4 dr - \int_0^{r_n} b_1^{large} r dr\right)$$

$$= \sum_{k,l} C_{k,i}^{sL^*} C_{l,j}^{pL} N_k^{sL^*} N_l^{pL} \left(-\frac{2\pi}{5\sqrt{3}} r_n{}^2\right) \tag{26}$$

$$\left[F_2{}^{small}(r_n)\right]_{i,j}$$

$$= \sum_{k,l} C_{k,i}^{sS^*} C_{l,j}^{pS} N_k^{sS^*} N_l^{pS} (\langle s|\boldsymbol{n}|p\rangle)_z \left(\frac{1}{r_n^3}\int_0^{r_n} b_1^{small} r^4 dr - \int_0^{r_n} b_1^{small} r dr\right)$$

$$= \sum_{k,l} C_{k,i}^{sS^*} C_{l,j}^{pS} N_k^{sS^*} N_l^{pS} \left(\frac{12\pi}{5\sqrt{3}} \alpha_k^* r_n{}^2\right) \tag{27}$$

(2) The cases for the combinations $\left(s_{1/2}^{+1/2}, p_{3/2}^{+1/2}\right)$, $\left(s_{1/2}^{-1/2}, p_{3/2}^{-1/2}\right)$, $\left(p_{3/2}^{+1/2}, s_{1/2}^{+1/2}\right)$, and $\left(p_{3/2}^{-1/2}, s_{1/2}^{-1/2}\right)$**.** For the $(p, s)$ combinations, the expressions can be obtained by interchanging the roles of the orbitals in the corresponding $(s, p)$ cases.

$$\left[F_1{}^{large}(r_n)\right]_{i,j} = \sum_{k,l} C_{k,i}^{sL^*} C_{l,j}^{pL} N_k^{sL^*} N_l^{pL} (\langle s|\boldsymbol{n}|p\rangle)_z \int_0^{r_n} b_1^{large} r dr = \sum_{k,l} C_{k,i}^{sL^*} C_{l,j}^{pL} N_k^{sL^*} N_l^{pL} \frac{2\sqrt{2}\pi}{3\sqrt{3}} r_n{}^2 \quad (28)$$

$$\left[F_1{}^{small}(r_n)\right]_{i,j} = \sum_{k,l} C_{k,i}^{sS^*} C_{l,j}^{pS} N_k^{sS^*} N_l^{pS} (\langle s|\boldsymbol{n}|p\rangle)_z \left(\int_0^{r_n} b_1^{small} r dr\right) = 0 \quad (29)$$

$$\left[F_2{}^{large}(r_n)\right]_{i,j} = \sum_{k,l} C_{k,i}^{sL^*} C_{l,j}^{pL} N_k^{sL^*} N_l^{pL} (\langle s|\boldsymbol{n}|p\rangle)_z \left(\frac{1}{r_n^3} \int_0^{r_n} b_1^{large} r^4 dr - \int_0^{r_n} b_1^{large} r dr\right)$$

$$= \sum_{k,l} C_{k,i}^{sL^*} C_{l,j}^{pL} N_k^{sL^*} N_l^{pL} \left(-\frac{2\sqrt{2}\pi}{5\sqrt{3}} r_n{}^2\right) \quad (30)$$

$$\left[F_2{}^{small}(r_n)\right]_{i,j} = \sum_{k,l} C_{k,i}^{sS^*} C_{l,j}^{pS} N_k^{sS^*} N_l^{pS} (\langle s|\boldsymbol{n}|p\rangle)_z \left(\frac{1}{r_n^3} \int_0^{r_n} b_1^{small} r^4 dr - \int_0^{r_n} b_1^{small} r dr\right) = 0 \quad (31)$$

***2. Electronic terms of analytical representation***

(1) The cases for the combinations $\left(s_{1/2}^{+1/2}, p_{1/2}^{+1/2}\right)$, $\left(s_{1/2}^{-1/2}, p_{1/2}^{-1/2}\right)$, $\left(p_{1/2}^{+1/2}, s_{1/2}^{+1/2}\right)$, and $\left(p_{1/2}^{-1/2}, s_{1/2}^{-1/2}\right)$. For the (*p*, *s*) combinations, the expressions can be obtained by interchanging the roles of the orbitals in the corresponding (*s*, *p*) cases.

$$\left[F_1{}^{large}(r_n)\right]_{i,j} = \sum_{k,l} C_{k,i}^{sL^*} C_{l,j}^{pL} N_k^{sL^*} N_l^{pL} \left(\frac{1-e^{-\alpha'_{k,l} r_n{}^2}}{2\alpha'_{k,l}}\right) \frac{4\pi}{3\sqrt{3}} \tag{32}$$

$$\left[F_1{}^{small}(r_n)\right]_{i,j} = \sum_{k,l} C_{k,i}^{sS^*} C_{l,j}^{pS} N_k^{sS^*} N_l^{pS} \left(\frac{\alpha_k\left(-3\alpha'_{k,l}+2\beta_l+e^{-\alpha'_{k,l} r_n{}^2}(-2\beta_l+\alpha'_{k,l})(3-2\beta_l r_n{}^2)\right)}{2\alpha'_{k,l}}\right) \frac{4\pi}{3\sqrt{3}} \tag{33}$$

$$\left[F_2{}^{large}(r_n)\right]_{i,j} = \sum_{k,l} C_{k,i}^{sL^*} C_{l,j}^{pL} N_k^{sL^*} N_l^{pL} \left(-\frac{3e^{-\alpha'_{k,l} r_n{}^2} r_n + 2\alpha'_{k,l} r_n{}^3}{4\alpha'_{k,l}{}^2 r_n{}^3} - \frac{3\sqrt{\pi}\, erf\left[\sqrt{\alpha'_{k,l}}\, r_n\right]}{8\alpha'_{k,l}{}^{5/2} r_n{}^3}\right) \frac{4\pi}{3\sqrt{3}} \tag{34}$$

$$\left[F_2{}^{small}(r_n)\right]_{i,j} = \sum_{k,l} C_{k,i}^{sS^*} C_{l,j}^{pS} N_k^{sS^*} N_l^{pS} \left(\frac{\alpha_k(3\alpha'_{k,l}-2\beta_l)}{\alpha'_{k,l}{}^2} - \frac{3\alpha_k e^{-\alpha'_{k,l} r_n{}^2}\left(5\beta_l+\alpha'_{k,l}(-3+2\beta_l r_n{}^2)\right)}{2\alpha'_{k,l}{}^3 r_n{}^2} + \frac{3\alpha_k(-3\alpha'_{k,l}+5\beta_l)\sqrt{\pi}\, erf\left[\sqrt{\alpha'_{k,l}}\, r_n\right]}{4\alpha'_{k,l}{}^{7/2} r_n{}^3}\right) \frac{4\pi}{3\sqrt{3}} \tag{35}$$

(2) The cases for the combinations $\left(s_{1/2}^{+1/2}, p_{3/2}^{+1/2}\right)$, $\left(s_{1/2}^{-1/2}, p_{3/2}^{-1/2}\right)$, $\left(p_{3/2}^{+1/2}, s_{1/2}^{+1/2}\right)$, and $\left(p_{3/2}^{-1/2}, s_{1/2}^{-1/2}\right)$. For the ($p$, $s$) combinations, the expressions can be obtained by interchanging the roles of the orbitals in the corresponding ($s$, $p$) cases.

$$\left[F_1{}^{large}(r_n)\right]_{i,j} = \sum_{k,l} C_{k,i}^{sL^*} C_{l,j}^{pL} N_k^{sL^*} N_l^{pL} \left(\frac{1-e^{-\alpha_{k,l}' r_n{}^2}}{2\alpha_{k,l}'}\right)\frac{4\sqrt{2}\pi}{3\sqrt{3}} \tag{36}$$

$$\left[F_1{}^{small}(r_n)\right]_{i,j} = \sum_{k,l} C_{k,i}^{sS^*} C_{l,j}^{pS} N_k^{sS^*} N_l^{pS} \left(\frac{\alpha_k\beta_l\left(1-e^{-\alpha_{k,l}' r_n{}^2}\left(1+\alpha_{k,l}' r_n{}^2\right)\right)}{\alpha_{k,l}'{}^2}\right)\frac{4\sqrt{2}\pi}{3\sqrt{3}} \tag{37}$$

$$\left[F_2{}^{large}(r_n)\right]_{i,j} = \sum_{k,l} C_{k,i}^{sL^*} C_{l,j}^{pL} N_k^{sL^*} N_l^{pL} \left(-\frac{3\mathrm{e}^{-\alpha_{k,l}' r_n{}^2} r_n + 2\alpha_{k,l}' r_n{}^3}{4\alpha_{k,l}'{}^2 r_n{}^3} - \frac{3\sqrt{\pi} erf\left[\sqrt{\alpha_{k,l}'} r_n\right]}{8\alpha_{k,l}'{}^{5/2} r_n{}^3}\right)\frac{4\sqrt{2}\pi}{3\sqrt{3}} \tag{38}$$

$$\left[F_2{}^{small}(r_n)\right]_{i,j} = \sum_{k,l} C_{k,i}^{sS^*} C_{l,j}^{pS} N_k^{sS^*} N_l^{pS} \cdot$$

$$\left[-\frac{2\alpha_k\beta_l}{\alpha_{kl}'{}^2} - \frac{3\alpha_k\beta_l e^{-\alpha' r_n{}^2}}{\alpha_{kl}'{}^2} - \frac{15\alpha_k\beta_l e^{-\alpha' r_n{}^2}}{2\alpha_{kl}'{}^3 r_n{}^2} + \frac{15\alpha_k\beta_l\sqrt{\pi} Erf\left[\sqrt{\alpha_{kl}'} r_n\right]}{4\alpha_{kl}'{}^{7/2} r_n{}^3}\right] \cdot \frac{4\sqrt{2}\pi}{3\sqrt{3}} \tag{39}$$

**Acknowledgements**

The authors would like to express their gratitude to Prof. N. Yoshinaga, Prof. B. P. Das, Prof. A. Vutha, and Dr. N. Yamanaka, for their insightful discussions on NSM physics. We also thank Mr. T. Fujita for his preliminary NSM calculations and Prof. K. Okada for his contributions during discussions in our laboratory. The authors further thank the Research Center for Computational Science (RCCS)/Institute for Molecular Science (IMS) in Okazaki, Japan, for providing access to their high-performance computer for large-scale calculations (Projects: 23-IMS-C049 and 24-IMS-C049).

**Disclosure statement**

The authors report there are no competing interests to declare.

**Fundings**

This work was supported by the Japan Society for the Promotion of Science (JSPS) KAKENHI Grant Numbers JP23K25893, and JP22K14031, and the research grant program of the Tokyo Ohka Foundation for the Promotion of Science and Technology.

**Data availability statement**

The data supporting the findings of this study are available within the article and its Supplementary Information.